\documentclass{article}

\title{\sc Local, Deterministic\\Hidden Variable Theories\\Based on a Loophole
\\in Bell's Theorem}

\author{Vladimir Z. Nuri \\ \small{ {\tt <vznuri@netcom.com>}} }
\date {June 1998}

\usepackage{graphicx}
\usepackage{psfrag}

\begin{document}
\maketitle
\begin{quote}
\footnotesize
{\bf Abstract:} This paper furthers the long historical 
examination of and debate on the foundations
of quantum mechanics (QM) by presenting 
two local hidden variable (LHV)
rules in the context of the EPRB experiment
which violate Bell's inequality, but which are nevertheless 
local and deterministic under reasonable definitions of the terms, and
coincide approximately with the conventional QM
prediction. The theories
are based on the general idea of probabilistic 
detection of particles depending on an interaction
of hidden variables within the measuring device and particle, and
relate mathematically to Fourier analysis.
The crucial discrepancy of variations in the
hidden variable distribution based on relative polarizer orientations is
isolated
which invalidates assumptions
in Bell-type theorems.
The first
theory can be analyzed completely symbolically whereas 
the second was analyzed using numerical methods. The properties of the
second in particular are shown to be approximately consistent with
the reported results and uncertainties in 
all three Aspect experiments. Variation in the total photon pairs
detected over orientations is shown to be a basic characteristic
of these theories. Some comments on the relevance of active vs.\ passive
locality are made. Two sections consider these ideas relative
to energy conservation and the measurement problem (collapse of the
wavefunction). One section proposes
new experiments.
\end{quote}
\section{Introduction}
The famous debate about the completeness of quantum mechanics (QM) 
as a physical theory dates at least to the publication of Einstein's `EPR'
paper in 1935~\cite{epr}, with earlier seeds in 
the Solvay physics conferences where Einstein
debated Bohr via carefully contrived {\em gedanken} experiments.
Almost seven decades of intense debate, including major theoretical
contributions by Bohm~\cite{bohm} and Bell~\cite{bell}, 
and experimental ones by 
Aspect~\cite{aspect1,aspect2,aspect3},
have clarified but not resolved the issue in the minds of physicists.
Wick~\cite{wick} is a good informal exposition focusing 
on historical details of 
the Bell inequality and Aspect experiments, and Baggott~\cite{baggott}
does so with more emphasis on the mathematical formalism. Greenstein
and Zajonc~\cite{greenstein} focus on the paradoxes of QM
with useful derivations. I found these sources very useful for the
multilayer analysis (ranging from the level of mathematical formalism to
historical background) that follows.

A dissatisfaction and skepticism for the 
adequacy of the Copenhagen interpretation 
promoted by Bohr has emerged in recent times, 
observed by e.g.\ Bernstein~\cite{bernstein} pp.~38-46.
Entire books have been dedicated to the topic of reexamination of
the foundations of QM, 
particularly relative to the EPR experiment, such 
those of Wick and Greenstein. A consensus has emerged that
the foundations of QM are exceptionally solid relative
to the history of scientific theories but
that there might nevertheless be 
some deeper order yet waiting to be uncovered. 
This theme has been explored aggressively by 
Bohm~\cite{bohm2}. The proscription of the
Copenhagen interpretation is, to paraphrase 
informally, ``there is no deeper reality.''
The intensity with which hidden variable theories have been pursued,
particularly in recent decades, indicate the lack of widespread
confidence and adherence to this ideology. The interpretation, taken
to an extreme, can become the equivalent of an 
unscientific self-fulfilling prophecy: nothing further will be 
found if nothing further is sought.

Wick in particular describes how a continual series of modifications and
exceptions have
been made to the conclusions of Bell-type analysis of the EPRB experiment,
based on the feedback between theoreticians and experimenters,
in a sort of continuous ongoing 
game of ``find the loophole.'' The search has been motivated by 
so-called ``confused realists'' looking for ways of preserving their
philosophy despite baffling experimental results. 
After long study myself
of the aformentioned excellent expositions  and some careful tinkering
I have solidified what I consider an excellent candidate for
a {\em bona fide} theoretical breakthrough, which is at least worthy of
further exploration. 

Within the 
fruitful context and framework of the EPRB correlated
particle problem, I've found two variations on a hidden 
variable theory that is local and deterministic
while still predicting the same results 
of QM (those experimentally observed by Aspect).  These
are extensions of some empirical results simultaneously considered
and observed by David Elm (see the acknowledgements in the 
final section).
This paper will describe this theory, first dissecting and revealing
its consistency and plausibility, particularly relative to the careful
and established results of Aspect, with a later section devoted
to some further speculation on its properties.
\section{Example LHV theory}
Greenstein and Zajonc have a nice mathematical analysis and
derivation of a sample local hidden
variable (LHV) theory in chapter 5~\cite{greenstein}, pp.~119-122. 
In what follows I will
borrow heavily from their presentation.  A very similar version
can be found in Baggott~\cite{baggott} (pp.~125-130). Each
reference also derives Bell's theorem in this context. In what
follows I will presume the reader is familiar 
with the basic EPRB setup and Bell's derivation. 
Merely to set up
a framework that can be utilized similarly for 
the new theory, I will show
the process of derivation for this familiar `naive' or `toy' 
hidden variable theory sometimes considered in the literature (the first
case I am aware of is in~\cite{baggott}).

\begin{figure}
\centering
\psfrag{th}[Bl][Bl]{$\theta$}
\psfrag{ph}[Bl][Bl]{$\phi$}
\psfrag{a}[Bc][Bc]{$\hat{a}$}
\psfrag{b}[Bc][Bc]{$\hat{b}$}
\psfrag{Vb}[Br][Br]{$V_b=-V_a$}
\psfrag{Va}[Bl][Bl]{$V_a$}
\includegraphics[scale=.5]{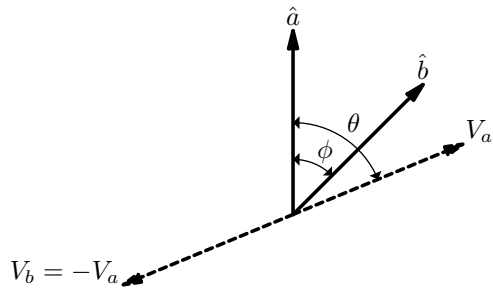}
\caption{Simple hidden variable theory of spin for the EPRB experiment}
\label{vectors}
\end{figure}

An elementary example of a hidden variable theory to describe
the EPRB experiment based
on an objective reality might be as follows. Two Stern-Gerlach
analyzers at opposite ends of the source are aligned relative
to directions specified by the vectors $\hat{a}$ and $\hat{b}$.
The `particle' is ejected from the source with a ``spin direction'' 
corresponding to the $V_a$ vector on the $\hat{a}$ detector side and
$V_b$ for the $\hat{b}$ detector. In the anticorrelated spin source
case, $V_b = -V_a$ (see Figure~\ref{vectors}). These vectors $V_a$ and
$V_b$ are the ``hidden variables'' that precisely determine 
the eventual spin measurements.
The detectors will respond either ``spin up'' or ``spin down''
as defined by two quantities $A$ and $B$ equal 
to $+1$ or $-1$ respectively. The product $AB$ of the two 
random variables (in the sense of being statistical quantities) 
will have an expected value over a 
large number of experiments as predicted by QM, 
$E_{\rm{QM}}$. According to the theory, 
\begin{equation}
E_{\rm{QM}}(\hat{a}, \hat{b}) = -\cos(\phi)
\label{qm}
\end{equation}
where $\phi$ is the angle of separation between the two detectors, 
i.e.\ the angle between vectors $\hat{a}$ and $\hat{b}$.

As Greenstein and Zajonc observe, 
a notable, basic, and natural parallel of the hidden
variable theory exists to an earlier era of scientific inquiry in 
which it was proven fundamental and essential.
The existence of atoms was not completely definitive during 
the development
of statistical mechanics laws for the behavior of gases. 
Scientists of the era conjectured the existence of hidden variables
such as atomic velocities to explain macroscopic observable quantities
such as pressure, and the tremendously successful accuracy 
of the ensuing theories was taken
as strong evidence in favor of the atomic hypothesis. Hence, pressure
properties (measurable) can be derived from the overall sum or average 
distribution of atom velocities (hidden). 
The statistical nature
of the model is then seen and realized 
as not intrinsic to the underlying phenomena
but related to lack of knowledge of hidden variables,
the crucial explanation that is instead denied 
by the Copenhagen interpretation
in the case of QM. In fact, even Planck's breakthrough
of quantifying energy transfer could be considered a successful early 
hidden-variable theory.

Analogously to the technique above, to 
compute the expected value of $AB$ relative to the sample theory,
it is necessary to calculate the product $AB$ for each possible value
of the hidden variable and then average over all possible values to
obtain the result, 
here labelled $E_{\rm{HV}}(\hat{a}, \hat{b})$. Assume the
hidden variable is randomly distributed over all orientations during
measurements, and
that the detector actually registers the sign of the projection of
the spin vector along its axis.
From the figure, $A = \rm{sign}[\cos(\theta)]$ and $B 
= \rm{sign}[\cos(180 + \theta - \phi)]
= \rm{sign}[-\cos(\theta - \phi)]$; hence $AB = 
\rm{sign}[-\cos(\theta)\cos(\theta - \phi)]$. 
The average (expected) value of $AB$ over all 
possible values of the hidden
variable $\theta$ is:
\begin{equation}
E_{\rm{HV}}(\hat{a},\hat{b}) = \frac{1}{2\pi}\int_{-\pi/2}^{3\pi/2} AB d\theta
\end{equation}
The integral is simple to compute because the product $AB$ takes on
exclusively the constant values $+1$ or $-1$ over the four-part 
interval as indicated in 
Table~\ref{funcab}. The end formula, which is not consistent with QM,
is not relevant to
any further considerations here but 
since following derivations have parallels to the calculation
I will include it here:
\begin{table}
\caption{The function $AB(\theta)$}
\label{funcab}
\[
\begin{array}{lcl|cc|l}
\theta_n & & \theta_{n+1} & A & B & AB \\
\hline
-\frac{\pi}{2}        & \rm{to} & -\frac{\pi}{2}+\phi & + & + & +1 \\
-\frac{\pi}{2} + \phi & \rm{to} & \frac{\pi}{2}        & + & - & -1 \\
\frac{\pi}{2}          & \rm{to} & \frac{\pi}{2} + \phi & - & - & +1 \\
\frac{\pi}{2} + \phi   & \rm{to} & \frac{3\pi}{2}      & - & + & -1 \\
\end{array}
\]
\end{table}
\begin{equation}
E_{\rm{HV}}(\hat{a},\hat{b}) = \frac{1}{2\pi}
\left\{ 
\begin{array}{l}
(+1)  \left[ \left( -\frac{\pi}{2} + \phi \right) - \left( -\frac{\pi}{2} \right)  \right] \\
+(-1) \left[ \left( \frac{\pi}{2} \right) - \left( -\frac{\pi}{2} + \phi \right) \right] \\
+(+1) \left[ \left( \frac{\pi}{2} + \phi \right) - \left( \frac{\pi}{2} \right) \right] \\
+(-1) \left[ \left( \frac{3\pi}{2} \right) - \left( \frac{\pi}{2} + \phi \right) \right] \\
\end{array}
\right\} = \frac{2}{\pi}\phi - 1
\label{naive}
\end{equation}
\section{Novel LHV theories overview}
This section introduces 
the underlying framework of new theories in this paper. In general, 
I will be presenting two LHV theories that are based on a probabilistic
detection of a particle at each detector as a function of a hidden
variable. So, consider a function $f(\lambda')$ where $\lambda'$ is
the hidden variable. This function will model the key aspects of the
hidden variable theory as follows. 

Typically the LHV theories in the literature consider a hidden variable 
$\lambda$, given in `absolute' coordinates, as the parameter to
the spin detection functions 
$f_a(\hat{a},\lambda), f_b(\hat{b},\lambda)$ where $\hat{a},
\hat{b}$ are the absolute orientations of the analyzers. 
Yet in contrast one can assume a local coordinate
(angle $\lambda'$) of the hidden variable
relative to each detector orientation without loss of generality. 
In the example
in the previous section, $\theta$ and $\theta - \phi$ are the angles of
the hidden variable relative to detectors $A$ and $B$ respectively. This
will allow the major simplification of looking 
for a single detection function 
$f(\lambda')$ common to each detector such that the 
expectation value of the correlation associated with the detection function
gives the predictions made by QM. In this formulation
$f(\lambda')$ is different in that not only is it given in local coordinates,
but also in that it gives the {\em probability} 
of detection for a given value of $\lambda'$ during $dt$,
i.e.\ analogous to a probability {\em density} function,
such that integration is required to give the cumulative probability of 
detection.

Again without loss of generality, assume the range of the 
hidden variable is $0 \leq \lambda' \leq 2\pi$. For symmetry purposes
let the function be periodic over this interval. Now as alluded, 
consider a case where $f(\lambda')$ does not directly give the detected 
spin orientation at a given detector depending on whether (say, for detector
$A$ with orientation $\hat{a}$)
a function $f_a(\hat{a},\lambda)$ is $\pm 1$, the 
basic and common formulation considered in the literature to date
by e.g.~\cite{greenstein} p.~123 and originally by Bell~\cite{bell}. 
Instead, the following scheme will be used.
$| f(\lambda') |$, i.e.\ the absolute value of $f(\lambda')$, will give
the {\em probability} the particle is detected; hence $| f(\lambda') | \leq 1$. 
If $f(\lambda') > 0$, the
particle will be ``spin up'' if detected; similarly the
``spin down'' case is implied by $f(\lambda') < 0$.

The prediction of QM as derived by Clauser et.\ al.\
in their seminal 1969 paper~\cite{clauser1}
for a photon-based version of Bell's experiment
is that the number of particles detected simultaneously at each detector
should be constant over different orientations of the analyzers. Or, 
equivalently, the rate of photons detected per time from a steady source
should be constant. This rate is given by the 
variable $R_0$ in their paper which will be considered very closely later
herein.
For the $f(\lambda')$ formulation considered here, however,
the condition implies the following. 

If $\theta$ is the hidden variable,
then $| f(\theta) f(\theta-\phi) |$ gives the probability a particle is
detected simultaneously by both detectors, i.e.\ a pair is measured at
some instant $dt$. 
Integrating the probability density function
over all hidden variable values yields the {\em cumulative} probability function,
which should be constant, say $t$, for 
all relative orientations of the detectors $\phi$:
\begin{equation}
t(\phi) = \int_0^{2\pi}| f(\theta)f(\theta-\phi) | d\theta = t
\label{total}
\end{equation}
 However, I
will be considering versions of $f(\lambda')$ 
for which it is only true that
$t(\phi) \approx t$.

The expected value of the correlation function can be computed as follows.
One can view the function $f(\lambda')$ as the product of the
spin value $\pm 1$ with 
the absolute probability of detection. Then
\begin{equation}
c(\phi) = \int_0^{2\pi} f(\theta)f(\theta-\phi) d\theta
\label{corrn}
\end{equation}
must give the expected correlation function for all $\phi$, the 
product of detected
spins over all $\theta$, 
not normalized to the number of pairs detected. The expected
value of the correlation function is normalized to the total pairs
detected, $t(\phi)$ in~\ref{total},
i.e.\ $E_{\rm{HV}'} = c(\phi)/t(\phi)$.
The prediction of
QM for spin-anticorrelated pairs~\cite{greenstein} p.~118
is that $E_{\rm{HV}'} = -\cos(\phi)$. One can summarize all of the above by
writing
\begin{equation}
E_{\rm{HV}'} = \frac{c(\phi)}{t(\phi)} =
\frac{\int_0^{2\pi} f(\theta)f(\theta-\phi) d\theta}
{\int_0^{2\pi}| f(\theta)f(\theta-\phi) | d\theta} = -\cos(\phi).
\label{key}
\end{equation}
with the condition that the denominator $t(\phi)$ 
is approximately equal to some
constant $t$ over all orientations $\phi$.
\section{Fourier transform relevance}
The integrals in Eq.~\ref{key} have interesting mathematical properties;
they can be computed using the theory of Fourier transforms. 
Let $F(u) = {\cal F}\{f(x)\}$ denote the Fourier transform 
operator on a function $f(x)$. The {\em convolution} of two
functions $f(x)$ and $g(x)$, denoted by $f(x) * g(x)$, is defined
by the integral
\begin{equation}
f(x) * g(x) = \int_{-\infty}^{\infty}f(\alpha)g(x-\alpha)d\alpha
\end{equation}
Define the {\em autoconvolution} of a 
function $f(x)$ as $f(x) * f(x)$.
Then the numerator and denominator of Eq.~\ref{key} are autoconvolutions,
and Fourier theory states that if $f(x)$ has the Fourier transform
$F(u) = {\cal F}\{f(x)\}$, then $f(x) * f(x)$ has the Fourier transform
$F(u)^2$. This can be written
\begin{equation}
f(x) * f(x) \Leftrightarrow F(u)^2
\end{equation}
In other words, to compute the numerator of Eq.~\ref{key}, 
one can take the
Fourier transform of the function $f(\lambda')$, square it, and then take the
inverse Fourier transform:
\begin{equation}
c(\phi) = {\cal F}^{-1}\{ {\cal F} \{ f(\lambda') \}^2 \}
\end{equation}
The denominator can be computed analogously using an $f'(\lambda')=|f(\lambda')|$.
I will not give a mathematical proof of these properties,  which follow
straightforwardly from basic Fourier theory,
but I have verified these properties numerically for the two hidden variable
theories considered herein. The technique relies on the periodicity of
the function assumed above and it being `even', 
i.e. $f(\lambda')=f(-\lambda')$.
\section{Preliminary critique}
The two theories in this paper do not {\em exactly} reproduce the 
predictions of QM. They differ slightly in ({\em a}) 
the correlation
expectation curve $E_{\rm{HV}'} \approx E_{\rm{QM}}$ (Eqs.~\ref{key}, \ref{qm}), and 
do not predict ({\em b}) a strictly unvarying rate of 
particle pairs detected over all orientations, the function $t(\phi) \approx t$ 
(Eq.~\ref{total}). I do not take `theory~{\sc i}' below seriously because of the
very large variation in ({\em b}), the total particles detected over different
orientations, even though it is conceivably consistent with the 
first Aspect experiment~\cite{aspect1}. It is interesting for its 
tractable and analytical
mathematical properties (results can be expressed in closed form) that
demonstrate potential implications of these types of theories, and
furthermore is very similar to the second theory~{\sc ii}.

However for `theory~{\sc ii}', as discussed below there is
very near agreement for both ({\em a}) and ({\em b}) to QM 
predictions, possibly even with the
experimental uncertainty bounds of all three Aspect experiments, including
the latter two~\cite{aspect2,aspect3} that measured total pairs detected. 
Disregarding experimental error, however,
admittedly the prospects for perfect theoretical agreement 
on {\em both} ({\em a}) and ({\em b})
via further tinkering with these types of theories (namely those based on
probabilistic detection of the particles based on the hidden variable)
are doubtful. 

Specifically regarding ({\em b}), I suspect there is 
a simple proof that $t(\phi)$ cannot be constant. Consider the following
informal argument. As noted above, the denominator $t(\phi)$ of 
Eq.~\ref{key} represents an `autoconvolution'. Hence the problem reduces
to finding a function for which its autoconvolution is a constant,
$t(\phi)=t$. Or in other words, any solution is among the functions $g(\lambda')$
such that $t(\phi) = {\cal F}^{-1}\{{\cal F}\{g(\lambda')\}^2\} = t$ 
(from the definition of autoconvolution). Using the algebraic properties 
of Fourier transforms, inverting this gives the unique function
$g(\lambda') = {\cal F}\{{\cal F}^{-1}\{t\}^{1/2}\}$. But any sequence of
Fourier transforms or inverses of a constant must be a constant. This 
would correspond to a situation in which the probability of detection of
a particle doesn't vary for $\lambda'$, precisely the case under study.

Regarding ({\em a}), it is not clear to me that there is no variation
of this type of hidden variable theory that will not give  
$E_{\rm{HV}} = E_{\rm{QM}}$ exactly even though $t(\phi) \approx t$. 
I have empirically tested many functions $f(\lambda')$ numerically 
and haven't found anything that
can be `tweaked' to arbitrary close degrees of agreement, but neither have
I found a proof that no such functions exist, although such a proof might
be tricky. Theory~{\sc ii} below is the
best compromise in simultaneously satisfying
({\em a}) and ({\em b}) I have found to date.

Clearly for these theories there is some relationship between the 
nearness of approximations of $t(\phi)$ and $E_{\rm{HV}}(\phi)$ to
QM predictions. I do not know whether 
a `better' theory than {\sc ii} is possible, in the sense that
a different detection function could yield even better
approximations.
However using the
Fourier theory techniques mentioned earlier I speculate it would be 
possible to derive an expression for the tradeoff between these two
parameters that would give a definitive answer. I want to emphasize
that even if it can be shown that the Aspect data is unequivocally
incompatible with theory~{\sc ii}, a new variation based on similar
principles as {\sc i,ii} (i.e. probabilistic detection) probably cannot 
necessarily be ruled out without such a ``tradeoff formula.''
\section{The loophole}
Finding any function $f(\lambda')$ such 
that Eq.~\ref{key} holds would
give a LHV for the aspects of QM embodied in the 
particle-pair 
spin-correlation experiments, and hence it is of paramount importance. 
It is not immediately obvious that
Bell's theorem applies to this situation, in 
which there is a variable 
detection probability at each detector 
dependent on the hidden variable, although the theory appears local
so far. 

I will argue that this theory can be both deterministic, local, and not
covered by Bell's theorem (that is, it nevertheless violates his inequality
in close accord with QM)
based on a remarkable loophole not previously
considered by other researchers. I am not suggesting that Bell's analysis
is mistaken for the cases he considers, only that there is a very
tricky subtlety in his mathematical definition of `locality' that 
is unexpected but can be cleverly exploited.
Bell's assumption in his theorem is
that there exists a hidden variable $\lambda$ that precisely determines
the spin detected at each detector. Even if there is only probabilistic
particle detection as I consider, it would seem one can consider the 
subset of cases in which both particles are detected.  Then it appears
to me relevance of
the theorem hinges on what determines whether the particle is detected 
as follows.

Consider the case where there is some hidden property within 
the particle, $\lambda_d,$ that determines whether it is registered by
either detector. In other words it has a domain of values denoted by the
set $\lambda_d \in D$, for which there are subsets 
$D_a \subset D$ and $D_b \subset D$ 
that overlap, $D_{ab} = D_a \cap D_b \neq \emptyset$, which comprises 
the pairs for which both particles are detected. 
Then there is some subset of its hidden values, namely $D_{ab}$, for which
one can apply Bell's theorem utilizing new
functions $f_a(\hat{a},m_a(\hat{a}-\lambda)), 
f_b(\hat{b},m_b(\hat{b}-\lambda))$ 
that are exclusively $\pm 1$ 
over the domain $\lambda_d \in D_{ab}$, where $m_a(\lambda'),
m_b(\lambda')$ are {\em local} mapping functions.  In this case
$m_a(\lambda')=\hat{a}-\lambda', m_b(\lambda')=\hat{b}-\lambda'$,
and one can successfully convert or reduce to the 
standard Bell forms $f_a(\hat{a},\lambda), f_b(\hat{b},\lambda)$.

Next, consider the case where there are hidden variables $\lambda_a,
\lambda_b$ {\em within each detector} that determine whether 
the particles are detected. If the distribution of $\lambda$
stays the same over all orientations even in this case, then
Bell's theorem applies as well.

Now consider the case where hidden variables in each detector
$\lambda_a,\lambda_b$ {\em in conjunction} with $\lambda$ determine
whether particles are detected. In other words, deterministic 
detection functions with values $\pm 1$ are
$f_a(\hat{a},\lambda_a,\lambda)$ and $f_b(\hat{b},\lambda_b,\lambda)$,
which are consistent with Eq.~\ref{key}.
Even though this dependence can result from exclusively local interactions,
in this situation Bell's theorem is apparently {\em not} applicable.
Evidently there exist 
no {\em local} mapping functions $m_a(\lambda_a,\lambda)=
m_b(\lambda_b,\lambda)$ to obtain Bell equivalence via a single
hidden variable with an identical distribution over orientations 
except in trivial cases in which the distributions of
$\lambda_a,\lambda_b$ are independent of $\lambda$.

Another way of stating this
problem is that the Bell theorem assumes a distribution for
the hidden variable $\lambda$ that does not vary over polarizer
orientations. Greenstein~\cite{greenstein} p.~123 says ``Bell's
theorem can be extended to the case of a nonuniform distribution
of $\lambda$s'' (without citation or proof;
Bell~\cite{bell}, p.~106 asserts the same). 
My clarification and
qualification is that it can be 
nonuniform, but must be an identical nonuniform distribution over 
all orientations. If each
detector contains hidden variables $\lambda_a,\lambda_b$ 
that, in {\em combination} with
the particle hidden variable property $\lambda$, determine whether
the particle is detected, the distribution of $\lambda$ over the
particle pairs detected can {\em vary} for
different polarizer orientations, even though there 
are no {\em nonlocal} physical events. In Bell's original paper~\cite{bell}
this limitation is in the form of the distribution function 
$\rho(\lambda)$, assumed dependent on a particle-centric
hidden property $\lambda$ only, p.~15. 

I have arrived at the previous ideas but do not have rigorous formal
proof for the {\em exact} reason Bell's theorem is violated and therefore
inapplicable in case of probabilistic particle detection
based on the hidden variable, although
the above sketch might be transformed into one by others.
Nevertheless the following LHV theory descriptions will provide numerical
proof of the violation of one simple version of Bell's theorem commonly
cited (see Greenstein~\cite{greenstein} p.~122, Bell~\cite{bell} pp.~18,38):
\begin{equation}
|E_{\rm{HV}}(\hat{a},\hat{b}) - E_{\rm{HV}}(\hat{a},\hat{c})| 
  \leq 1 + E_{\rm{HV}}(\hat{b},\hat{c})
\label{bell}
\end{equation}
One might argue the Bell theorem is not applicable in the following
theories not due to a varying hidden variable distribution but instead
simply because of the approximations ({\em a}) 
$E_{\rm{HV}'}(\phi) \approx E_{\rm{QM}}$ and/or
({\em b}) $t(\phi) \approx t$.  
The logical interplay and intertwined
dependency of the implications involved are subtle and in my opinion 
worth further
close scrutiny by others. 
For example, do ({\em a}) and/or ({\em b}) imply a varying
hidden variable distribution?  It might seem the whole question
is mute if it can be shown
that the class of theories based on detection as a function of
the hidden variable cannot be consistent with either theoretical or
experimental physical results, but I would argue their study would
still shed light on the precise scope of Bell's theorems. 
A section immediately following the mathematical details
critiques their plausibility, particularly relative to the Aspect
experiments.
\section{Hidden variable theory~{\sc i}}
\begin{figure}
\centering
\psfrag{mpi2}[Bc][Bc]{$-\frac{\pi}{2}$}
\psfrag{ph1}[Bc][Bc]{$-\frac{\pi}{2}+\phi$}
\psfrag{ph2}[Bc][Bc]{$\frac{\phi}{2}$}
\psfrag{ppi2}[Bc][Bc]{$\frac{\pi}{2}$}
\scalebox{.5}{\includegraphics{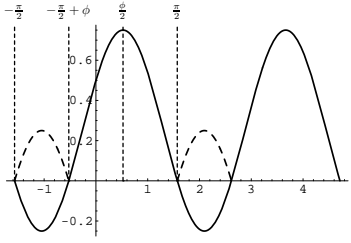}}
\caption{LHV theory {\sc i} 
$c'_1(\theta,\phi)$ and 
$|c'_1(\theta,\phi)|$ for $\phi=\frac{\pi}{3}$ (Eq.~\ref{f1})}
\label{extrema}
\end{figure}

This section will consider a more sophisticated hidden variable
theory that actually produces effects that could be consistent
with the first Aspect experiment,~\cite{aspect1}.  Consider
a simple detection function 
based on
the projection of the hidden variable $\lambda$ with the 
detector axis, and the corresponding correlation and total
count functions (Eqs.~\ref{corrn}, \ref{total}).
\begin{eqnarray}
f_1(\lambda') & = & \cos(\lambda') \label{f1} \\
c_1(\phi) & = & \textstyle{\int_{0}^{2\pi}} \cos(\theta) \cos(\theta-\phi) d\theta \label{c1}\\
t_1(\phi) & = & \textstyle{\int_{0}^{2\pi}} |\cos(\theta) \cos(\theta-\phi)| d\theta \label{t1}\\
c'_1(\theta,\phi) & = & \cos(\theta,\phi) \cos(\theta-\phi) \label{c1p}
\end{eqnarray}
The sign of $c'_(\theta,\phi)$ (Eq.~\ref{c1p})
gives the expected correlation product of
spins detected, $-1$ for opposite spins and $+1$ for the same. The
absolute value $|c'_1(\theta,\phi)|$ gives the probability of detecting
the pair. These two functions are graphed in Fig.~\ref{extrema}. From
the figure, it is clear the distribution of $\lambda$ (i.e. $\theta$)
changes over polarizer orientations $\phi$. 
The function $|c'_1(\theta,\phi)|$. 
is periodic on the interval
$[-\frac{\pi}{2},\frac{\pi}{2}]$, 
with minima of zero at those points and
$\theta=\pi/2+\phi$, where no pairs are detected; the maxima of one is at
$\frac{\phi}{2}$ where the pair is always detected as correlated.

Eq.~\ref{t1} can be integrated symbolically even with the absolute value
complication using the trick of 
separation into the intervals 
$[-\frac{\pi}{2},-\frac{\pi}{2}+\phi]$
where $c'_1(\theta,\phi)$ is negative and 
$[-\frac{\pi}{2}+\phi,\frac{\pi}{2}]$ 
where it is positive, 
and doubling the value of the integral for the full interval 
$[-\frac{\pi}{2},\frac{\pi}{2}]$ 
based on the periodicity.
The separation in fact is equivalent to dealing with the anticorrelated
and correlated cases separately, respectively.
In other words,  let the functions $c_-(\phi),c_+(\phi)$ represent
the probability $0 \leq c_{\pm}(\phi) \leq 1$ of detecting an anticorrelated 
or correlated pair relative to $\phi$, respectively, over the half-period. 
Then
\begin{equation}
\begin{array}{rcl}
c_-(\phi) & = & \textstyle{-\int_{-\pi/2}^{-\pi/2+\phi}} c'_1(\theta,\phi)d\theta \\
c_+(\phi) & = & \textstyle{\int_{-\pi/2+\phi}^{\pi/2}} c'_1(\theta,\phi)d\theta \\
t_1(\phi)/2 & = &
\textstyle{\int_{-\pi/2}^{-\pi/2+\phi}} |c'_1(\theta,\phi)|d\theta
+ \textstyle{\int_{-\pi/2+\phi}^{\pi/2}} |c'_1(\theta,\phi)|d\theta \\
& = & c_-(\phi)+c_+(\phi)
\end{array}
\label{correq}
\end{equation}
Symbolic integration and simplification via Mathematica software yields
\begin{equation}
\begin{array}{rcl}
2 c_-(\phi) & = & \sin(\phi) -\phi \cos(\phi) \\
2 c_+(\phi) & = & \sin(\phi) + (\pi - \phi) \cos(\phi) \\
t_1(\phi) & = & 2 \sin(\phi) + (\pi - 2\phi) \cos(\phi) 
\end{array}
\label{f1soln}
\end{equation}
\begin{figure}
\centering
\psfrag{a1}[Br][Br]{$c_+(\phi)$}
\psfrag{a2}[Bl][Bl]{$c_-(\phi)$}
\psfrag{a3}[Bc][Bc]{$t(\phi)/2$}
\scalebox{.5}{\includegraphics{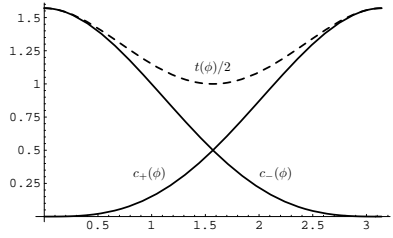}}
\caption{LHV theory {\sc i} functions 
$c_{\pm}(\phi)$ and $t_1(\phi)/2$ (Eqs.~\ref{correq})}
\label{corr}
\end{figure}

A graph of the two functions $c_{\pm}(\phi)$ and their sum $t_1(\phi)/2$
appears in Fig.~\ref{corr}.   The
total pair count curve has both upward and downward concavity.
The maximum deviation from the 
mean of $\frac{\pi}{4}+\frac{1}{2} = 1.2853$ 
is $\pm (\frac{\pi}{4}-\frac{1}{2}) = \pm .2853$ or $\pm 22.3\%$, with a 
standard deviation of $15.7\%$ for an equidistant 50 point sample over
the interval---clearly 
completely physically incorrect. However,
it does show exactly how a purely local theory can exhibit fluctuations in the
total photon pairs detected over polarizer orientations, even with
a constant flux into each detector during the interval $dt$. 

Conveniently,
$c_1(\phi)/2 = c_+(\phi) - c_-(\phi)$, so $c_1(\phi) = -\pi \cos(\phi)$.
Hence for this theory (Eq.~\ref{key}),
\begin{equation}
E'_{\rm{HV}} = \frac{c_1(\phi)}{t_1(\phi)} = - \frac{\pi \cos(\phi)}
{2 \sin(\phi) + (\pi - 2\phi) \cos(\phi)}
\label{hv1}
\end{equation}
\begin{figure}
\centering
\psfrag{ehv1}[Br][Br]{$E'_{\rm{HV}}$}
\psfrag{cos}[ul][ul]{$E_{\rm{QM}}=-\cos(\phi)$}
\psfrag{dhv1}[Bl][Bl]{$E'_{\rm{HV}} - E_{\rm{QM}}$}
\scalebox{.5}{\includegraphics{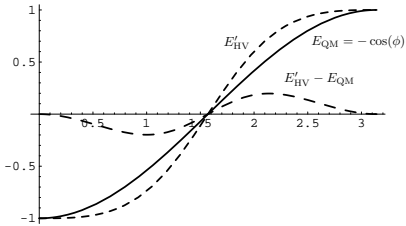}}
\caption{LHV theory {\sc i} prediction
$E'_{\rm{HV}}$ (Eq.~\ref{hv1}) vs. $E_{\rm{QM}}$ (Eq.~\ref{qm})}
\label{cmp}
\end{figure}
A comparison graph of $E'_{\rm{HV}}(\phi)$ vs. $E_{\rm{QM}}(\phi)$ 
is given in Fig.~\ref{cmp}.  The maximum absolute difference is 
$\pm 19.8\%$ with a standard deviation of $12.7\%$ for an equidistant
50 point sample over the interval. 
The graph is striking in that it 
reproduces the concavity of the QM prediction,
in contrast to the earlier naive hidden variable theory (Eq.~\ref{naive})
which predicts a straight line.

The Bell formula inequalities can now be computed and the
violation demonstrated 
based on Eq.~\ref{bell}, say for
$a=0,b=\frac{\pi}{3},c=\frac{2\pi}{3}$ 
with $E'_{\rm{HV}}(a,b) = E'_{\rm{HV}}(b-a)$:
\begin{equation}
\begin{array}{rcr@{.}l}
|E'_{\rm{HV}}(0,\frac{\pi}{3}) - E'_{\rm{HV}}(0,\frac{2\pi}{3})| & = & 1 & 39277 \\
1+E'_{\rm{HV}}(\frac{\pi}{3},\frac{2\pi}{3} & = & 0 & 30362
\label{belleq1}
\end{array}
\end{equation}
\section{Hidden variable theory~{\sc ii}}
A slight variation on theory {\sc i} above gives significantly
improved results. This section will consider a model for the case 
where {\em correlated} spin particles emanate from a souce, with
a reduction of the theory to the anticorrelated case requiring only
some simple geometric transformations (namely translation and scaling of
the hidden variable). Consider the function
\begin{equation}
f_2(\lambda') = \cos(\lambda')^{|1/e|}
\label{f2}
\end{equation}
where $e$ is the base of natural logarithms, $e \approx 2.7183$, and
$1/e \approx .3679$, and the definition of exponentiation 
in this function is adjusted
for negative values in the following way:
\begin{equation}
a^{|b|} = \left\{
\begin{array}{ll}
a^b & \textrm{if $a \geq 0$} \\
-(-a)^b & \textrm{if $a < 0$}
\end{array}
\right.
\end{equation}
\begin{figure}
\centering
\psfrag{f1}[Br][Br]{$\cos(\lambda')$}
\psfrag{f2}[Bl][Bl]{$\cos(\lambda')^{|1/e|}$}
\scalebox{.5}{\includegraphics{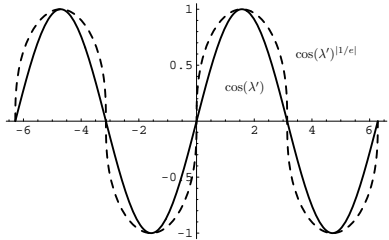}}
\caption{LHV theory {\sc i} $f_1(\lambda')$ (Eq.~\ref{f1}) vs. 
theory {\sc ii} $f_2(\lambda')$ (Eq.~\ref{f2})}
\label{f1f2}
\end{figure}
A graph of $f_2(\lambda')$ vs. $f_1(\lambda')$ (Eq.~\ref{f1})
can be found in Fig.~\ref{f1f2}. The unnormalized correlation (Eq.~\ref{corrn})
and total flux functions (Eq.~\ref{total})
are defined (analogously to Eqs.~\ref{c1}, \ref{t1}) as
\begin{eqnarray}
c_2(\phi) & = & \textstyle{\int_{0}^{2\pi}} \cos(\theta)^{|1/e|} \cos(\theta-\phi)^{|1/e|} d\theta \label{c2}\\
t_2(\phi) & = & \textstyle{\int_{0}^{2\pi}} |\cos(\theta)^{|1/e|} \cos(\theta-\phi)^{|1/e|}| d\theta \label{t2}
\end{eqnarray}
(again with the specialized exponentiation operator).

These integrals almost certainly have no closed-form solution yielding
a precise symbolic integration. However numerical integration is not
difficult. I wrote a simple program in Mathematica that used
rectangular integration over $50$ equally-spaced points for
$0 \leq \theta \leq \pi$ and the half-interval 
$0 \leq \phi \leq \pi$ (utilizing symmetry). 
To test it I applied it to
formula $f_1(\lambda')=\cos(\lambda')$ (Eq.~\ref{f1}) and 
found an identical result to
Eqs.~\ref{f1soln} with negligible discrepancy due to 
numerical inaccuracy.

\begin{figure}
\centering
\psfrag{t2}[ul][ul]{$t_2(\phi)/2$}
\psfrag{ehv2}[Bl][Bl]{$E''_{\rm{HV}} \approx E'_{\rm{QM}}= \cos(\phi)$}
\scalebox{.5}{\includegraphics{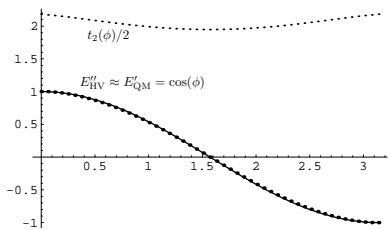}}
\caption{LHV theory {\sc ii} $t_2(\phi)/2$, $E''_{\rm{HV}}$ vs. $E'_{\rm{QM}}$}
\label{f2cmp}
\end{figure}

\begin{figure}
\centering
\psfrag{dt}[ul][ul]{$\frac{t_2-\overline{t_2}}{\overline{t_2}}$}
\psfrag{dhv2}[Br][Br]{$E''_{\rm{HV}} - E'_{\rm{QM}}$}
\scalebox{.5}{\includegraphics{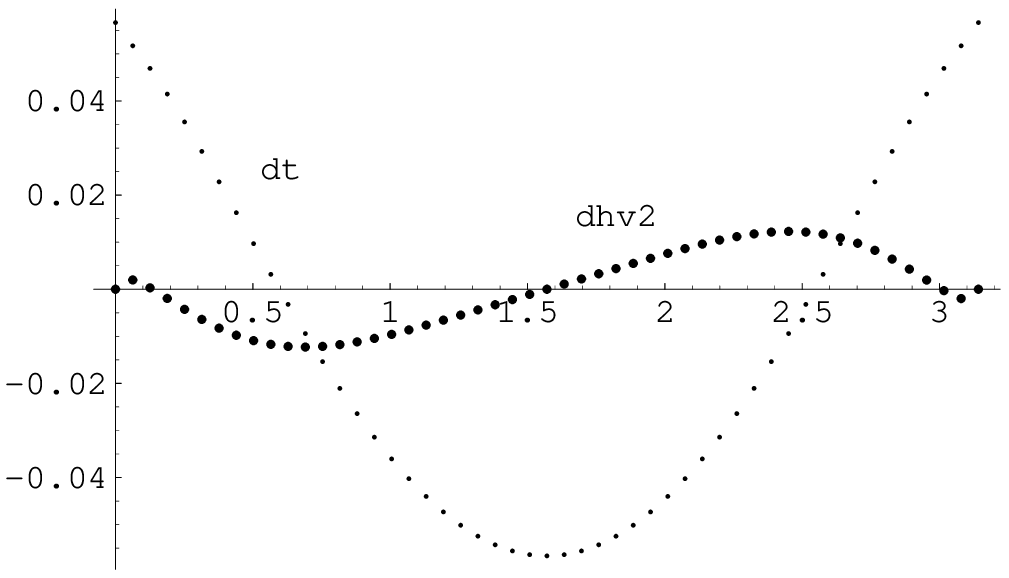}}
\caption{LHV theory {\sc ii} differences 
$(t_2 - \overline{t_2})/\overline{t_2}$ and $E''_{\rm{HV}}- E'_{\rm{QM}}$}
\label{f2diff}
\end{figure}

The results of the integration and normalization
$E''_{\rm{HV}}(\phi) = c_2(\phi)/t_2(\phi)$
for Eqs.~\ref{c2}, \ref{t2} are shown in Fig.~\ref{f2cmp} with
$t_2(\phi)/2,E'_{\rm{QM}}$ in the same graph. For correlated particles,
$E'_{\rm{QM}}(\phi) = \cos(\phi)$ where here $\phi$ 
is twice the angular difference
between polarizer orientations 
(see e.g. Greenstein~\cite{greenstein}, p.~136).
More impressively for this theory~{\sc ii}, the total particle pairs
detected is closer to a constant (the mean $\overline{t_2}/2 = 2.07$). 
Most remarkable is the very near
equivalence to the QM prediction $E'_{\rm{QM}}$. The differences
in these two key measures are shown in Fig.~\ref{f2diff}. $E''_{\rm{HV}}$
differs from $E'_{\rm{QM}}$ by a maximum of $\pm 1.2\%$ with a standard 
deviation of $.80\%$ (sampled over the 50 points). 
$t_2$ differs from the mean $\overline{t_2}$ by
a maximum of $\pm 5.7\%$ with a standard deviation of $3.7\%$. In contrast
to the first theory (Fig.~\ref{corr})
the total pair detection count curve appears to be concave 
upward only.

Because of the close correspondence between $E''_{\rm{HV}},E'_{\rm{QM}}$ the 
theory~{\sc ii} function $f_2$ violates the Bell inequality almost
exactly in accordance with QM.
\section{Experimental plausibility}
The meticulous experiments of Aspect~\cite{aspect1,aspect2,aspect3} are
extremely useful and crucial constraints for investigating the
plausibility of any alternative QM theory, particularly
any supposedly LHV varieties. The experiments are widely regarded as
definitive proof in favor of conventional QM, or at
least the impossibility of any local hidden variable theories. Nevertheless
the
results above suggest a review of the exact experimental findings
of the papers would be reasonable.

Consider the first Aspect experiment,~\cite{aspect1}. In this experiment,
because of a polarizer setup that either blocked or passed photons of
a particular polarization,
assumptions had to be made about the equivalence of the absence of detected 
photons to the existence of photons of opposite polarization from those
detected. That is, the experiment could not detect
photons of both `horizontally' and `vertically' 
oriented spins at each detector. The effect of this
is that the experiment cannot measure total photons detected over all
orientations. Aspect et.\ al. then must measure the maximum photon rate
detected for aligned polarizers, i.e. $R_0$, and assume it stays constant
over all orientations. 

But viewing Eq.~\ref{hv1} of theory {\sc i}, it
is apparent that the theory will predict correct results if the expectation
value is not normalized to the varying total number of photon pairs over all 
orientations. Eq.~\ref{f1soln} will give a $t_{\rm{max}} = \pi$ measured
in the aligned case, and $c_1(\phi)/t_{\rm{max}} = -\cos(\phi)$. I
regard this only as an interesting theoretical 
curiosity, however, given that
the later Aspect experiments indeed measured 
constant total photon pairs detected.

Nevertheless the second Aspect experiment~\cite{aspect2} had 
experimental uncertainties that are conceivably consistent
with theory~{\sc ii}. They reported a $\pm 1\%$ variation in
the detection efficiency of the polarizers and a variation
from $E'_{\rm{QM}}$ ``better than $1\%$.'' These two uncertainties
combined might be enough to be within the bounds of theory~{\sc ii}.

Most importantly Aspect et.\ al. conspicuously did not 
report the crucial uncertainty in the `constant' fourfold-sum coincident 
detection rate (or simply, the photon pair rate)
$R_{\pm\pm}(\vec a,\vec b)$, although they do
state the rate itself 
was `typically' 80 $s^{-1}$ (\cite{aspect2}, p.~93). I would
consider it striking if it
varied in a curve similar to that of Fig.~\ref{f2cmp}.

A final factor in these experiments that might support LHV theory~{\sc ii} 
is a theoretical distinction that has been asserted by
Santos in~\cite{santos}. He insists that the expression for the
``depolarization factor'' derived by Clauser et.\ al. in~\cite{clauser1}
is in fact dependent on the polarizer angle difference as well as
the detector light cone angle, $F(\theta,\varphi)$,
whereas their expression is only
dependent on the latter. (Note that Santos renamed some variables
rather confusingly relative to their paper and my own. In his
case $\theta$ is the polarizer angle difference and $\varphi$ is
the detector light cone half-angle. Clauser et.\ al.\ use $\theta$
to refer to the light cone half-angle and $\varphi$ to the polarizer
angle difference.) If there is fundamental
disagreement on this point about `depolarization' related to 
polarizer angles then it is conceivable that even theory~{\sc i} could be 
more than a theoretical curiosity.
\section{Active vs. passive locality}
Diverse researchers have formulated increasingly rigorous and
precise qualifications on Bell's theorem 
and isolated its key assumptions since its origin. 
Theories 
based on probabilistic detection 
dependent on hidden variables
are not specifically addressed in general but surely 
fall under known general categories, although their exact classification
is a delicate issue. Even though they are approximate they may raise issues
and reveal potentials.

An excellent technical analysis of Bell's theorem's precise
nature is given by Faris in an appendix to Wick~\cite{wick},
pp.~227-279. He frames the distinction observed in the previous
section in that ``QM
cannot be reduced to probability on a single probability model. \ldots
The first Bell theorem rules out the possibility that QM
is described by a unique probability measure.'' However, a `family'
of probability measures depending on the measurement ``indeed is possible''
(p.~270). He indicates that Bell was aware of the 
exception and formulated a ``second theorem'' 
intended to make more manifest these hidden assumptions, apparently
referring to~\cite{bell} pp.~105-110.

Faris makes the distinction between active and passive locality, the two 
crucial prerequisites for the new theorem. A violation
of active locality would support a nonlocal physics capable of
superluminal signalling. On the other hand, Faris writes on the
other condition,
``If deterministic passive
locality were violated, it would be possible that there would
be randomness at the measurement stage with no origin in any preparation
stage, yet which would still maintain the perfect correspondence between
events at distant locations'' (p.~272). (Determinism is
implied by passive locality, p.~244.)

The theories of this paper appear to me to violate `passive locality.'
If so
I object to misleadingly labelling the latter property, ``deterministic 
passive locality,'' either a `local' or `deterministic'
constraint, because the theories violate it (via the
probability of detection in 
the detectors) yet are fully local and deterministic. 
The `randomness' referred to by Faris 
may be that the probability of detection of particles is dependent
on an interaction of hidden variables within both the measuring device
and the particles. This would be not so much 
`nonlocal' as it is `nonreductionistic'
such that any model that takes only the device or the particle properties
into account separately must invoke counterintuitive
and unrealistic `nonlocal' effects. Wick~\cite{wick} writes p.~220,
\begin{quote}
The existence of the apparatus, it seems, transforms the hidden
variables describing the particle pair. Either these variables are
altered deterministically depending on the settings of {\em both}
analyzers (violating active locality on a hidden level) or Nature
throws dice in a way depending on both (violating passive locality).
\end{quote}
But theory~{\sc ii} of this paper exhibits
a varying {\em distribution} of
hidden variables over orientations even via a local 
and deterministic physics, yet
yields close agreement with QM predictions.
Faris writes on p.~278, ``a violation of passive locality would only
mean that dependence between simultaneous events at distant locations need
have no explanation in terms of prior events. This is not clearly ruled
out, but it is not evident how to construct such a theory.'' One might
qualify and rephrase
this as ``\ldots no explanation in terms of prior events {\em based
on a reductionistic perspective.}'' Eq.~\ref{key} is a mathematical
constraint sufficient 
for such a theory, and the LHV theories~{\sc i,ii} satisfy it
to varying degrees of approximation.

Others have isolated many of the subtle prerequisites of the Bell
theorems in theory and in practice. 
A contrived exception to exhibiting the Bell 
inequality in practice that depends on
detector efficiency upon which a LHV theory can
be constructed is noted in a comprehensive 1978 review
article by Clauser and Shimony (p.~1913). 
They write, ``Although the selection
is done locally, it does have the appearance of being highly artificial
and, indeed, almost conspirational against the experimenter.'' 
My findings are that 
generally the LHV theories~{\sc i,ii} are novel relative to prior
literature overall, and my opinion in contrast is that they are not 
at all physically implausible. 
\section{Beam density fluctuation}
Clauser and Shimony consider
various qualifications such as the `CHSH' and weaker `CH' assumptions
about the probabilities of detecting particles relative to polarizer
equipment,~\cite{clauser2} p.~1904. 
Santos~\cite{santos}
constructed a mini-theory for the aligned polarizer case
based on a violation of the latter CH ``no enhancement'' hypothesis
($P(\lambda,a) \leq 1$) 
but not related to detector efficiency.   He defines a 
photon density function in terms of two hidden variables 
$\rho(\lambda_1,\lambda_2)$, analogous to my own use of a probability 
density function, where $\lambda_1,\lambda_2$ are associated with the
separate detectors and detection, for say $a$, is proportional to the product
$\rho(\lambda_1,\lambda_2)P(\lambda_1,\hat{a})$, which 
he apparently presumes can be given in terms of a function of the detector
variable $\lambda_1$ and the particle variable $\lambda$.  However,
he doesn't explain this in particular, i.e. how such a density might 
come about based on exclusively local effects.

There is some further
discussion of $\rho(\lambda)$ on pp.~105-110 of Bell~\cite{bell}. He
insists that ``the difficulty would not arise\dots if $\rho_1$ were
allowed to depend on $b$, or $\rho_2$ on $a$. Such a dependence would
not only be of mysteriously long range, but also, for the case 
presented, would have to propagate faster than light. The correlations
of QM are not explicable in terms of local causes.'' I
disagree that a $\rho_1(\lambda,\lambda_a,\lambda_b)$ 
would necessarily be nonlocal,
with theory~{\sc ii} again as the counterexample. (Although this unexpected
dependence is certainly counterintuitively {\em nonreductionistic}.)
The paper seems to me to get caught up in the crucial distinction between
correlation and causality:
\begin{quote}
Now surely it would be very remarkable if the choice of program in Lille
proved to be a causal factor in Lyon, or if the choice of program in Lyon
proved to be a causal factor in Lille. It would be very remarkable, that
is to say, if $\rho_1$\ldots had to depend on $b$, or $\rho_2$ on $a$.
But, according to QM, situations presenting just such a 
dilemma can be contrived.
\end{quote}
But a function $\rho_1(\lambda,\lambda_a,\lambda_b)$ may embody a 
{\em correlation} or bias that is implicit in the selection sample, without
implying {\em causality}, even when it appears the chosen sample ought not to
have biased properties. This is apparently the ``passive locality'' 
assumption which as indicated might be more accurately termed
``passive {\em reductionism}.''

In general it appears to me prior research
has not explicitly isolated the crucial distinctions between 
detection functions that depend on detector-based hidden variables
interacting in a dependent way with the particle property $\lambda$,
although Santos seems to have unconsciously invoked it. 
The difficulty is aggravated in that very subtle formal
differences between equivalent and
inequivalent mathematical formulations do not make the
differences obvious.
\section{Energy conservation}
David Elm (see the final section) has collected some objections
to a theory based on a variation in detection depending on the
hidden variable. One main objection by his correspondents is that
such a theory would violate conservation of energy. Looking closer at this
claim, in general conservation of energy in physical theories 
can be stated in several different ways:
\begin{eqnarray}
E_{\rm{in}} & = & E_{\rm{out}}  \label{a}\\
E_{\rm{in}} & \geq & E_{\rm{out}}  \label{b}\\
E_{\rm{in}} & = & E_{\rm{out}} + s \label{c}
\end{eqnarray}
(\ref{a}) implies all energy is precisely accounted for. (\ref{b}) and
(\ref{c}) are equivalent, where the former suggests that energy is
dissipated in the system that isn't measured, and the latter implies
that it is contained in some internal property that cannot be directly
measured, say $s$, such as
entropy. A supposed ``violation of energy conservation'' with
novel LHV theories is
only true in the sense that (\ref{a}) is not applicable if
not all energy
can be accounted for because some particles are not always detected. But this
is not a physical impossibility, because (\ref{b}) or (\ref{c}) may
still apply, and indeed are routinely 
invoked to describe virtually all physical experiments. 

Generally I would say that a ``violation of energy conservation'' 
(in the sense of the laws of thermodynamics) only
really occurs if it can be shown that $E_{\rm{out}}>E_{\rm{in}}$, neither of
which (\ref{b}) or (\ref{c}) imply. That is, it is not at all necessary 
to account for all energy fluctuations exactly, except
unless the theory claims to be
{\em complete}, in which case the inequality
(\ref{b}) would indeed seem to be inadequate. 
But this would be a problem for QM, not any LHV theory
that implicitly insists that conventional QM is incomplete.
Actually, for this reason, any experiment that could show energy that was
consistently not accounted for by QM would tend to support
the existence of any other theory (\ref{c})
that somehow accounted for
unmeasurable dissipation `$s$' at least with theoretical satisfaction.

Theories~{\sc i,ii} indeed postulate unmeasured energy 
in the form of particles (photons)
released from the source but not detected.
An `independent'
calculation of energy in the experiment could possibly reveal 
the discrepancy.  However typically in experimental arrangements no
such information can be determined; the energy released by the source can 
only be inferred by the flux into detectors.
\section{Collapse of the wavefunction}
Finally, a short idea speculating on the measurement problem relative to 
theories~{\sc i,ii} is apropo here. The crucial {\em nonreductionistic}
but not {\em nonlocal} interaction between experimental apparatus and the 
measured object is revealed by these theories. 
In the particle-pair correlation 
experiments, hidden variables $\lambda_1,\lambda_2$ within each
detector might {\em interact} with the hidden variable $\lambda$ within the
particle to determine detection probabilities, even locally, and in
conjunction lead to, but not {\em cause}, a nonlocal {\em correlation}. 
(Equivalently, the distribution of values for the
particle-based hidden variable $\lambda$ varies depending on orientation for
the detected subset of pairs.)

Perhaps after this
interaction, the hidden variable $\lambda$ within the particle is
fixed to the measured state. This might be sufficient to explain the
notoriously mysterious ``collapse of the wavefunction'' that has
plagued QM thought since its inception. In other 
words, in the
language of decoherence (see e.g. Greenstein~\cite{greenstein} ch.\ 8),
prior to the measurement the state of the system
is a nonlocal {\em superposition} of
$\lambda_1,\lambda_2,\lambda$, but the measurement leads to a {\em mixture}
in which $\lambda$ `collapses' to some measured state {\em if detected}.

Somewhat related to this, 
some researchers are pursuing LHV theories based on a return to 
a locally realist theory of the electromagnetic wave with nonclassical
fluctuations of the Zero Point Field (ZPF)
 that would be regarded as `noise' in QM 
theories~\cite{marshall1,marshall2,marshall3}. 
I am partial toward these
efforts for which theory~{\sc ii} may be compatible.
\section{Proposals}
As stated, Aspect observed $\pm 1\%$ detection efficiency variations 
over orientations in the second
experiment~\cite{aspect2}, which might actually
be related to photon density variations as predicted by theory~{\sc ii},
Fig.~\ref{f2cmp}.  They also observed $\pm 1\%$
variation in $E_{\rm{expt}} \approx E_{\rm{QM}}$. Their combined {\em reported} 
uncertainties are approximately large enough such that the results
could be considered consistent with LHV theory~{\sc ii}.

Based on the ideas in this paper and the Aspect reports I have a 
few suggestions for physical experiments:
\begin{itemize}
\item The existing data for the Aspect experiments could be reexamined and
reinterpreted in light of these findings since they did not report
the crucial degree of variation in the `constant' fourfold-sum coincident 
detection rate (or simply, photon pair rate) $R_{\pm\pm}(\vec a,\vec b)$. 
In particular this key measurement of
total photon pair counts could be plotted over orientation. If any trend
is observed, it could be favorable toward theory~{\sc ii}.
\item Or, for photons, new experiments could try to gain greater accuracy
than the existing Aspect experiments 
which are widely considered definitive. Some would not consider this important,
but precision in other physical constants has been pursued at great effort 
to distinguish between competing theories,
such as the measurement of the magnetic moment of the electron to verify
the predictions of quantum electrodynamics (QED). Arguably the locality of
QM is at least as fundamental and paramount.
\item New Bell-type experiments that involve particles with masses instead
of massless photons might be performed. In some ways the similarity of
predictions of QM for these two seemingly widely divergent
situations is remarkable. Unfortunately experimental intractability is high.
\end{itemize}
\section{Acknowledgements}
Here I will give a brief footnote about the historical priority of some
of these
ideas for future reference.  In 1997 I speculated that
a rule based on a hidden variable that determined the probability of
a particle being detected might simultaneously explain the results
and exhibit both determinism and locality. Toward this end I found 
the excellent intuition and guidance of Wick's book particularly absorbing, 
and the straightfoward mathematical derivations of 
Greenstein very helpful. I intended to experiment
empirically with different formulas, starting with the simplest possible
I could imagine, in which the probability of detection was proportional
to the projection of the hidden variable direction vector onto the 
detector direction vector. 

In 1998, after an extensive search of the
web prior to any tinkering, 
I discovered David Elm's results (web site: 
{\tt www.tiac.net/users/davidelm/epr.htm};
email: {\tt <davidelm@tiac.net>})
which give empirical evidence
that the simple projection rule $f_1$ is
sufficient in some sense. However, at the time
of writing, Elm describes his results in somewhat simplistic and 
obscure terms of a game he calls ``Circles and Shadows.'' 

Elm apparently tested
a projection rule similar to $f_1$ using computer simulations, although I couldn't
verify this except in retrospect. The information
in his description is not presented in a mathematical format and
was somewhat unclear and not entirely sufficient for me to 
understand his formulation, and he did not include his 
code on the web site or respond to an email query on my part to obtain it.
Hence I couldn't verify he was referring to the simple projection rule
until I obtained similar results by somewhat independent analysis, and
seeing in retrospect our expectation curves (which he plotted) were 
approximately the same in each case. 

Elm frames his results as a demonstration of erroneous reasoning by Bell.
I prefer the point of view advanced above
that Bell was correct in his derivations but
a subtle loophole in his definitions of `local' and `deterministic'
can lead to a new theory, although the viewpoints are essentially 
identical. 

On the 
$R(\phi)/R_0$ issue, Elm writes: ``\ldots the `simple logic' used by Bell
seems to me to contain a vital flaw. I believe the mistake is manifested
in the way this reasoning causes one to scale of the data [{\em sic}] 
to plot it against the graph of the upper limit as 
put forth by Bell.'' He writes later:
\begin{quote}
It seems more logical that the detection of each photon arriving at each
polarizer is simply a probability based on the relative angles between the 
polarizer and the spin vector of the photon.
\end{quote}
Elm states this would imply
``a substantial amount of events would be missing from the data''
whereas ``Bell's inequality is only applicable if it includes 
all the data.'' However, Elm doesn't note that the second experiment of 
Aspect~\cite{aspect2} that detected constant total photon pairs for all 
polarizer orientations would tend to reject the 
simple theory in which total pairs
detected varies tremendously. In fact Elm's account is not aware 
at all of the key
constraint on any LHV theory verified 
by that experiment that total photon pairs detected
should be roughly constant. Hence I think
his arguments are mostly relevant (and limited) to the first Aspect
experiment in which the lack of measured events is equated 
with the existence of anticorrelated particle pairs, in spite of
his insistence that ``all EPR experiments have used this faulty
scaling as the basis of their determination of the validity, or lack of,
the local reality views\ldots''

In consideration of all this I would say that we have independently 
discovered at least
the implication of probabilistic yet local detection 
violating Bell's theorem, and possibly also
the same basic principle for the simple projection
rule $f_1$;
all justifiable credit goes to
Elm for his prior informal and empirical 
work in this area. I am grateful for his communication of the
knowledge and perspectives 
his experiments led to, which were explored, according to 
his web site, possibly as early as 1993. 
Various objections he posted to his web
site by respondents were also helpful in gauging the legitimacy of 
the theory and likely reactions, on the energy conservation discrepancy
in particular.
The mathematical analysis of the simple projection rule
$f_1$, the advanced projection rule $f_2$, and the following
observations and speculations are my own.

Finally, 
I would like to thank V.~S. for various resources and encouragement,
and Wolfram Research for the Mathematica software which was used
extensively throughout the paper.

\end {document}